\documentclass{elsart}
\usepackage{epsfig}

\begin{document}
\newcommand*\mycite[1]{\cite{#1}}
\newcommand*\mylabel[1]{\label{#1}}
\newcommand*\myref[1]{(\ref{#1})}

\begin{frontmatter}
\title{Volatility Cluster and Herding}
\author{Friedrich Wagner}
\address{Institut f\"ur Theoretische Physik,
Universit\"at Kiel, Leibnizstrasse 15, D-24098 Kiel, Germany}
\begin{abstract}
Stock markets can be characterized by fat tails in the volatility
distribution, clustering of volatilities and slow decay of their
time correlations. For an explanation
models with several mechanisms and consequently many parameters as the 
Lux-Marchesi model have been used. We show that a simple herding model with
only four parameters leads to a quantitative description of the data.
As a new type of data we describe the volatility cluster by  the waiting
time distribution, which can be used successfully
to distinguish between different models.
\end{abstract}
\end{frontmatter}

\section{Introduction} \label{intro}
Financial stock markets exhibit several universal properties, socalled
'stylized facts'\mycite{Vries,Pag}.
There are no time correlations in the index for time scales larger
than a few minutes \mycite{Fama,Stan2}. The distribution of relative
changes or returns are power law distributed at large values with an
exponent in the order of 3-4 \mycite{Stan2,Man}. The absolute returns or 
volatilities exhibit multifractal moments \mycite{Lob}. 
In the time development of the volatilities we find two features. The
correlation decays very slowly. Wether this decay is described by 
a power law with a small exponent or an exponential cannot be decided.
A power law would be equivalent with a $1/f$ noise \mycite{Stan2}. 
The Joseph effect \mycite{Man2}
manifests itself by volatility clusters, that large changes of the index
occur preferably at neighbouring times.\\
These facts are interesting enough to motivate also physicists to look
for microscopic models for an at least partial explanation. By microscopic
models we mean a market consisting of agents following certain trading
strategies together with a mechanism for the evolution of the price or 
the index. Many different approaches have been persued in the recent
years (for reviews see \mycite{Farm} or recently \mycite{Sam}). All
these models use inhomogenous agents in contrast to the effective
market hypothesis \mycite{Fama}. The agents are distinguished by
differences in the behaviour as noise traders or 
fundamentalists \mycite{KM,Bak,LM,MG},
the wealth \mycite{SLH}, the bidding prices \mycite{Sato},
the trading thresholds \mycite{Iori} or
the length of memory \mycite{LLS}. Alternatively equal agents select
different strategies depending on an utility function \mycite{Cal,CB,Stau,Bor}.
The mechanism for creating volatility clusters may be a memory, nonlinear
coupling between the price and agent parameters or the herding effect.
The latter may be acchieved by a next neighbour interaction as in
statistical mechanics \mycite{Iori,CB,Stau,Bor} or explicitely in the
dynamics \mycite{Bak,LM}. A model including all three mechanisms is the
Lux-Marchesi model \mycite{LM} (hereafter abbreviated by LMM). If such a
model accounts successfully for the stylized facts as it happens for LMM one
would like to know whether all three mechanisms are needed or one alone is
already sufficient. In the present paper we investigate the question
whether herding alone can explain the stylized facts.\\
As a model we use a simplified version of the LMM. 
As pointed out in \mycite{Bak}
there must be three type of agents for which we take noise traders
(pessimists and optimists) and fundamentalists as in LMM. An important
unique feature of the LMM is that agents can change their oppinion.
Herding is described by transition rates proportional to the number of agents
in the new state. If we neglect the dependence of the agent
numbers on the price we can
use the ant formalism of Kirman \mycite{Ker} or its generalization by 
Aoki \mycite{Aoki}. Since the price has disappeared from the dynamic we have to 
find a phenomenological relation in terms of the agent numbers.
For that we use the observation \mycite{AL}, that in the LMM the dynamics
of the price is much faster than that of the agents and therefore
the equilibrium relation can be used. Assuming a constant number of
fundamentalists the Kirman model can acchieve qualitative agreement
with the data \mycite{AL}. However, we will show that
the time structure of the clusters is in disagreement with the data. 
Moreover there exists a winning strategy for the 
fundamentalists, since noise traders are not allowed to become
fundamentalists. Therefore we use a three component herding model 
where herding is included also for the fundamentalists as in \mycite{Bak}.\\
The paper is organized in the following way. In the next section we
describe the data (German DAX index \mycite{DAX}). We use the volatility
distribution, the the correlation function of the volatility 
and as a new feature
the waiting times between volatilities larger than a certain minimum value.
Section \ref{price} is devoted to the description of the agent strategies
and the price relation. For large numbers of agents the Kirman model
can be written in terms of a Fokker-Planck equation. Instead of using
transition probabilities we formulate the dynamics by 
the corresponding Langevin equation for the fraction of agents. This model and
its consequences are
described in section \ref{2komp}. The formalism can be easily generalized
for three components (section \ref{three}). Our conclusions are
contained in section \ref{concl}.

\section {German DAX index} \label{data}
The indices $p(t)$ of stock markets
show a long term  increase. This is a rather small effect, f.e. for the 
German DAX index 0.03\%/day over 30 years.
Apart from that there are no time
correlations on time scales larger than a few minutes \mycite{Fama,Stan2}.
In contrast the absolute values of the return or the volatility defined by
\begin{equation} \mylabel{d1}
v(t)=\frac{1}{p(t)}\left | \frac{\Delta p}{\Delta t} \right |
\end{equation}
exhibits the stylized facts mentioned in section \ref{intro}, if we consider
dayly variations. Unfortunately also the values of $v(t)$ averaged over
few years show long term variations, as pointed out in \cite{Voit}. To avoid
a long term bias we chose the dayly quoted values of the DAX in the
time period 1996-1999 corresponding to 784 trading days, where the three month
average of $v(t)$ is compatible with a constant value $v_{av}=0.011$.
\begin{figure}[htb]
\let\picnaturalsize=N
\def\picsize{120mm}
\def\picfilename{fig1.ps}
\ifx\nopictures Y\else{\ifx\epsfloaded Y\else\input epsf \fi
\let\epsfloaded=Y
\centerline{\ifx\picnaturalsize N\epsfxsize \picsize\fi
\epsfbox{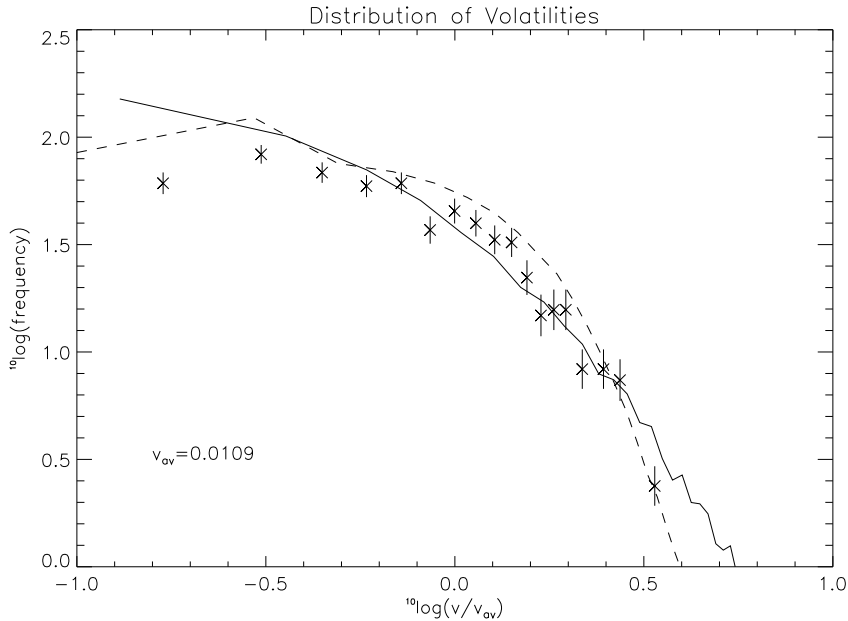}}}\fi
\medskip
\quad

\caption{\label{fig1} The distribution of normalized volatilities for the DAX
         is shown on a double log scale (data points). The solid (dashed) line
         show the prediction of the three component model (Lux-Marchesi model).}
\end{figure}
In Fig.\ref{fig1} we show the distribution of $v/v_{av}$ on a double 
log scale for these data. The use of normalized volatilities \cite{Stan2}
is important when comparing the data with models. Since $v_{av}$ defines
a time scale, the translation of computer time steps in a simulation 
into a real time is taken
into account using the normalized volatilities $v/v_{av}$. As one sees from  
Fig.\ref{fig1} both LMM and the three component model to be discussed in
section \ref{three} are compatible with the data. It is difficult to
decide whether there is a power law behaviour at large $v$ or not.
The errors of the data points shown in Fig.\ref{fig1} (and in the
subsequent figures) are computed by binning the volatilities and taking
$\sqrt{N_b}$ as statistical error, if the bin contains $N_b$ events.\\
We quantify the effect of volatility clustering by the waiting times
$\tau$ between occurences of volatilities larger than a minimum value
$v_{min}$ measuring the time $t=\tau \Delta t$ in units of $\Delta t=1$ day.
If the volatilities $v(\tau)$ are statistically independent the probability
$p_w(\tau)$ for a waiting time $\tau=1,2,...$ is given by
\begin{equation} \mylabel{d2}
p_w(\tau)=\frac{1}{\tau_0 -1}
          \left ( \frac{\tau_0-1}{\tau_0} \right )^{\tau} \quad ,
\end{equation}
where the average waiting time $\tau_0$ is the inverse of the cumulative
probability $P(v\ge v_{min})$.
\begin{figure}[htb]
\let\picnaturalsize=N
\def\picsize{150mm}
\def\picfilename{fig2.ps}
\ifx\nopictures Y\else{\ifx\epsfloaded Y\else\input epsf \fi
\let\epsfloaded=Y
\centerline{\ifx\picnaturalsize N\epsfxsize \picsize\fi
\epsfbox{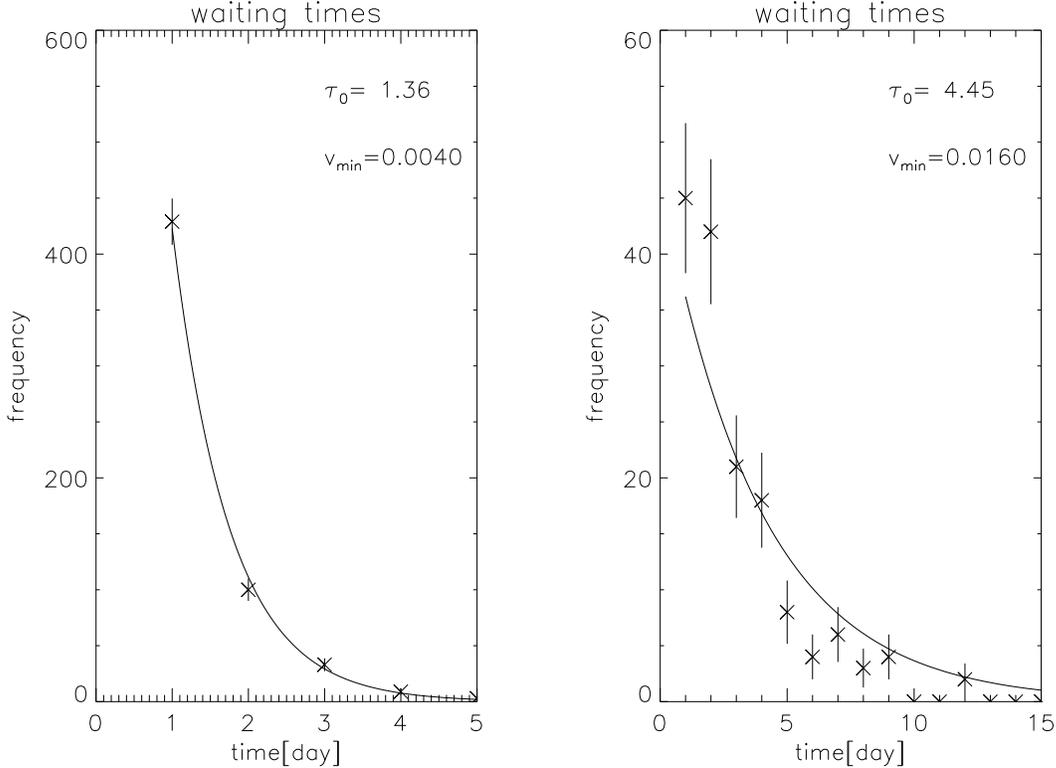}}}\fi
\medskip
\quad
\caption{\label{fig2} The distribution of the waiting time for a volatility
larger than $v_{min}$ for $v_{min}=0.004$ and 0.016. The curves correspond to 
independent volatilities and are given by equ.\myref{d2} using the 
observed average waiting time $\tau_0$.}
\end{figure}
In Fig.\ref{fig2} we show $p_w(\tau)$
for different $v_{min}$. For $v_{min}<<v_{av}$ it agrees with \myref{d2}
indicating that small changes of the index can be considered as independent.
For values $v_{min}\sim v_{av}$ we see an excess at small and large
value of $\tau$ and a depletion near $\tau \sim \tau_0$
as expected from formation of volatility clusters. The experimental
distribution can be characterized by the average waiting time $\tau_0(v_{min})$
and a clustering parameter $\gamma(v)$ defined by
\begin{equation} \mylabel{d3}
\gamma(v_{min})=\sum_{\tau =1}^{[2\tau_0]}\mbox{sign} (\tau_0-\tau)
     \left (\frac{N(\tau)}{N_0}-p_w(\tau)\right )
     /\sum_{\tau =1}^{[\tau_0]}p_w(\tau) \quad ,
\end{equation}
where $N(\tau)$ corresponds to the observed number of events at 
time $\tau$ and $N_0$
to the total number. The clustering parameter $\gamma$ is proportional to
the difference of the areas between the data and the curve $p_w(\tau)$
expected from independent volatilities in the intervalls $\tau <\tau_0$
and $\tau_0 <\tau <2\tau_0$. The average waiting time $\tau_0(v)$ is
\begin{figure}[htb]
\let\picnaturalsize=N
\def\picsize{140mm}
\def\picfilename{fig3.ps}
\ifx\nopictures Y\else{\ifx\epsfloaded Y\else\input epsf \fi
\let\epsfloaded=Y
\centerline{\ifx\picnaturalsize N\epsfxsize \picsize\fi
\epsfbox{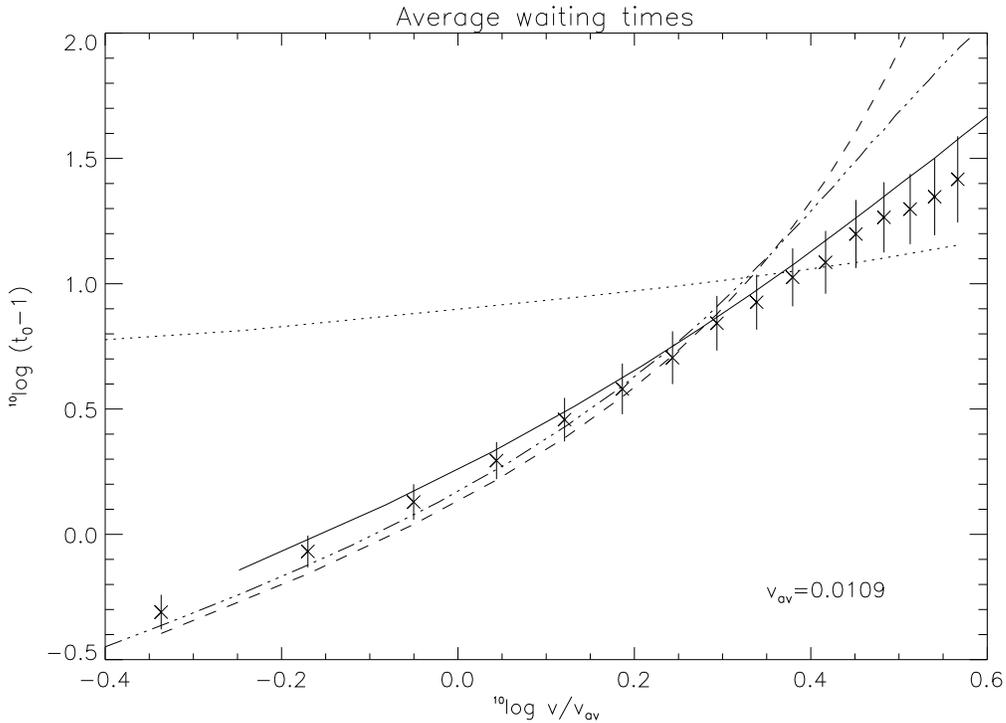}}}\fi
\medskip
\quad
\caption{\label{fig3} The average waiting time $\tau_0(v)$ for volatilities
larger than $v$ as function of the normalized volatility on a double log scale.
The solid (dotted,dashed,dotted-dashed) line show the prediction of the three
component model (Kirman model, random walk, Lux-Marchesi model).}
\end{figure}
compared in Fig.\ref{fig3} with independent Gaussean distributed
volatilities given by
\begin{equation} \mylabel{d4}
\frac{1}{\tau_0}=\frac{1}{\sqrt{\pi}}\; \mbox{erfc}(\frac{v}{v_{av}\sqrt{\pi}})
\end{equation}
and the LMM, which both are remarkably close to the data for
$v<2v_{av}$. Again we used the scaled values $v/v_{av}$ of the
minimum volatility $v=v_{min}$, which
allows an absolute prediction for one parameter models as in equ. \myref{d4}.
\begin{figure}[htb]
\let\picnaturalsize=N
\def\picsize{150mm}
\def\picfilename{fig4.ps}
\ifx\nopictures Y\else{\ifx\epsfloaded Y\else\input epsf \fi
\let\epsfloaded=Y
\centerline{\ifx\picnaturalsize N\epsfxsize \picsize\fi
\epsfbox{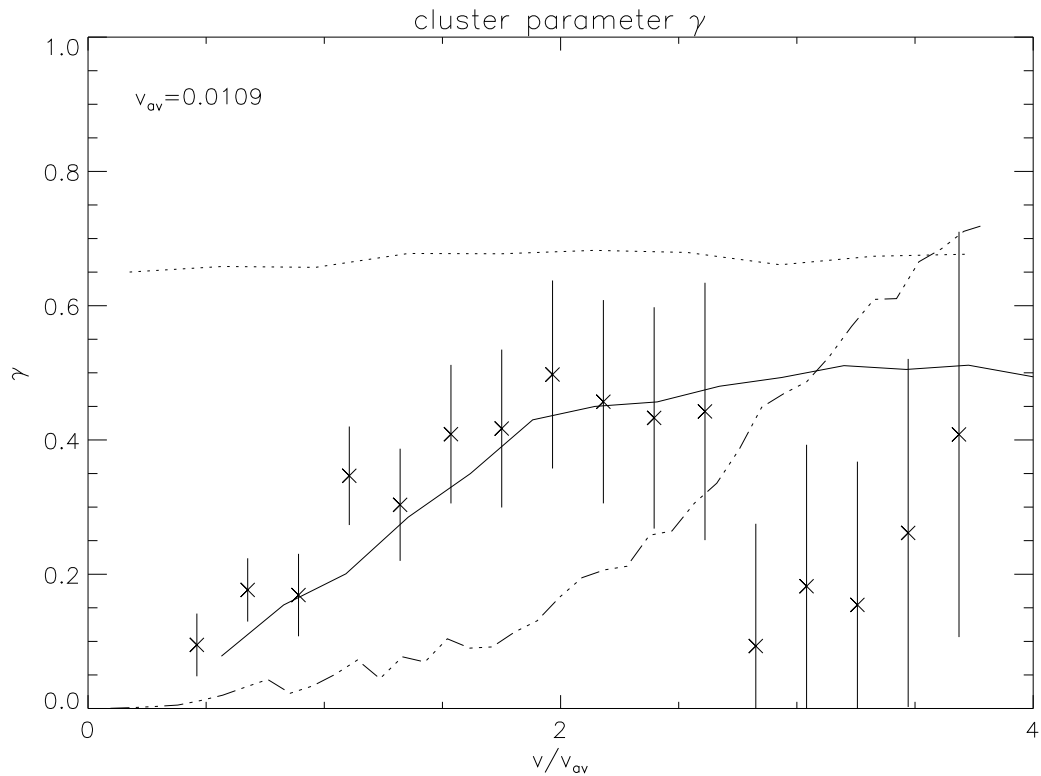}}}\fi
\medskip
\quad
\caption{\label{fig4} The data show the cluster parameter $\gamma$ defined
         by equ. \myref{d3} for the DAX as function of the normalized 
         volatility. The solid (dotted, dotted-dashed) line show the 
         prediction of the three component model (Kirman model,
         component model (Kirman model, Lux-Marchesi model).
         $\gamma=0$ corresponds to no clustering.}
\end{figure}
In Fig.\ref{fig4} the cluster parameter $\gamma(v)$ from
equation (\ref{d3}) is shown as function of $v/v_{min}$. The linear
increase of $\gamma(v_{min})$ shows that there exists clustering on all
scales of $v$. In the LMM and the Kirman model to be dicussed in
section \ref{2komp} the cluster are generated
by an intermittency effect by two almost stable states. In the second model
the fluctuations outside the cluster are almost zero and during the 
cluster of the same size leading to constant $\gamma$ and 
$\tau_0$ (see Fig.\ref{fig3}).
The reversed happens in the LMM. Here the fluctuations are independent for
$v<v_{av}$ and the
clustering begins only at large values of $v$. Note that the LMM has not
been adjusted to these data. In view of the many parameters in this 
model the disagreement should not to be taken too seriously.
As last experimental quantity we will use the correlation function defined by
\begin{equation} \mylabel{d5}
C(\tau)=\frac{\sum_{t=1}^{T}(v(t)v(t+\tau)-v_{av}^2)}
             {\sum_{t=1}^{T}(v(t)v(t)-v_{av}^2)} \quad .
\end{equation}
A meaningful experimental determination of $ C(\tau)$ requires much more
statistics than used in the previous figures. Therefore we use all DAX data
from 1969 to 2002. The correlation function for the volatility is shown
in Fig.\ref{fig5} on a log-linear scale. The data show a strong decrease
at small values of $\tau$ which compares well with the prediction of the LMM.
It fails to account for the values at large $\tau$, which in turn are in
agreement with the three component model.
\begin{figure}[htb]
\let\picnaturalsize=N
\def\picsize{130mm}
\def\picfilename{fig5a.ps}
\ifx\nopictures Y\else{\ifx\epsfloaded Y\else\input epsf \fi
\let\epsfloaded=Y
\centerline{\ifx\picnaturalsize N\epsfxsize \picsize\fi
\epsfbox{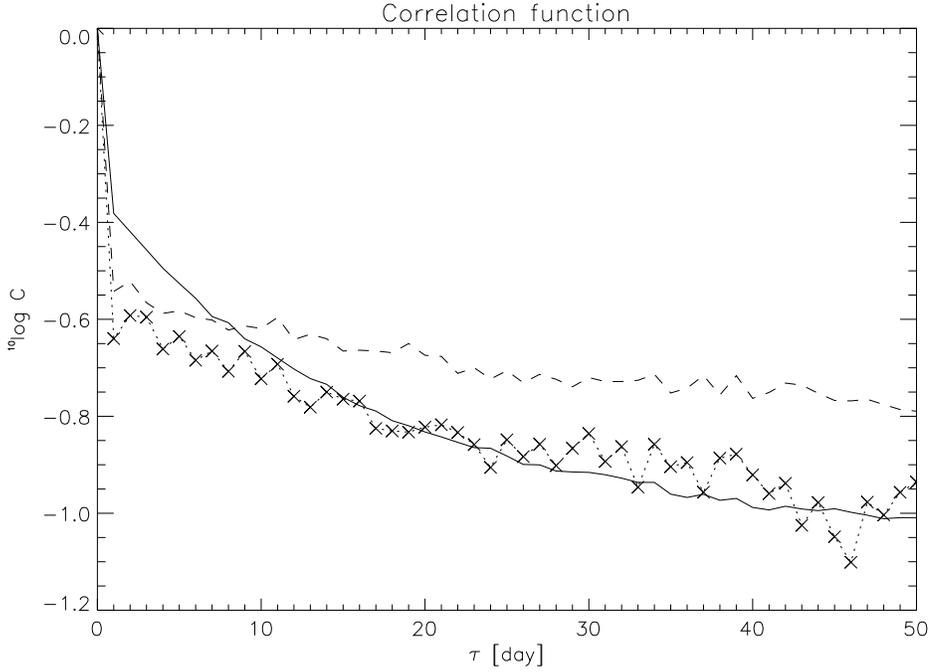}}}\fi
\medskip
\quad
\caption{\label{fig5} The correlation function for the DAX (crosses) on a
log-linear scale. An exponential decay would correspond to a straight line.
The solid (dashed) line gives the prediction of the three component model
(Lux-Marchesi model).}
\end{figure}

\section{Equation of state for the price} \label{price}
In order to explain the features of the DAX discussed in the
previous section \ref{price} we introduce as in the LMM\cite{LM} $N$
agents, which can adopt one of the following strategies. Optimistic
(pessimistic) noise traders buy (sell) at each time one unit of
stock. Fundamentalistic traders change their stock proportional
to the difference of the actual price $p(t)$ and an assumed 
true value $p_f$. The system is characterized by the numbers
$n_i(t)$ of agents following strategy $i=+,-f$ (optimistic,
pessimistic and fundamentalistic) and the price $p(t)$. In general
the dynamic will couple $n_i(t)$ and $p(t)$ in a complicated way.
The new price $p(t+\Delta t_p)$ will depend on the demand $D$ and
the supply $S$. To avoid dependence on splitting orders it should
depend only on the difference $\omega =D-S$ which is given in our
case by
\begin{equation} \mylabel{b01}
\omega (p)=\frac{n_f}{p_0}(p_f-p)\;+\;n_+ - n_- \quad .
\end{equation}
The following fairly general form of a price dynamic 
\begin{equation} \mylabel{b02}
p(t+\Delta t_p)=p(t)\; g\left (\frac{\lambda(t)}{\lambda(t)-\omega (p)}\right )
\end{equation}
has been discussed by Helbing et al.\mycite{Hel00}. 
In relation \myref{b02} $\lambda(t)$
can describe a reservoir of stocks to fulfill any order or the market
liquidity. The montonically increasing function $g$ must satisfy $g(1)=1$.
For the following it is irrelevant wether one assumes an exponential form
\mycite{Far98}, a linear function \mycite{LM} or a power law\mycite{Hel00}.
Motivated by the simulation of the LMM we assume the price changes due to
to the dynamic \myref{b02} occur at a much smaller time scale $\Delta t_p$
than changes of $n_i(t)$ or $\lambda(t)$ at the scale $\Delta t$. 
Therefore the dynamic \myref{b02} 
will reach a fixed point before any changes in $n_i(t)$ or $\lambda(t)$
occur. The mapping \myref{b02} has the stable fixed point $\omega (p)=0$
if the stability condition
\begin{equation} \mylabel{b03}
\left | 1-g'(1)\frac{p}{p_0}\frac{n_f}{\lambda}\right |\; < \; 1
\end{equation}
holds, which can be achieved by sufficient large liquidity $\lambda $.
From the fixed point condition $\omega (p)=0$ we obtain the equation of
state for the price
\begin{equation} \mylabel{b1}
p=p_f+p_0\;\frac{n_+-n_-}{n_f} \quad .
\end{equation}
This relation has been also proposed by Alfarano and Lux \mycite{AL}.
Alternatively one can consider (\ref{b1}) as a phenomenological relation
expressing that high (low) price are expected, if the optimists
(pessimists) and a price $p_f$ if fundamentalists
dominate the market. In order to avoid negative prices the second term
in relation (\ref{b1}) must be much smaller than $p_f$. Since the
average returns for the DAX or other markets are in the order of
1\% the probability for a negative price is very small. Neglecting
quadratic terms in $v$ we find for the volatility (\ref{d1})
\begin{equation} \mylabel{b2}
v(t)=\frac{p_0}{p_fn_f}\;\left |\frac{\Delta n_+}{\Delta t}-
     \frac{\Delta n_-}{\Delta t} \right | \quad .
\end{equation}

\section{Kirman model} \label{2komp}
In this section we simplify the model even further by keeping
the number of fundamentalists constant. This serves mainly to
explain the method. It can be easily generalized to three
components (see section \ref{three}). It has been introduced by
Kirman \mycite{Ker} for volatility clustering and behaviour of ants.
The noise traders follow the strategies denoted by $i=+,-$.
The transition rate $\pi$ per unit time for one agent to switch
from strategy $i$ to $k$ is given by
\begin{equation} \mylabel{c1}
\pi (i \to k)=a_{ik}+b_{ik}\cdot n_k \quad .
\end{equation}
The number of noise traders 
$n_c=n_++n_-$ is to be kept constant. The constants $a$
describe a spontaneous rate and $b$ correspond to the herding
behaviour. For symmetry between optimistic and pessimistic traders
$a$ and $b$ do not depend on $i,k$. The property $b_{ik}=b_{ki}$ has
the remarkable consequence \cite{Ker,Mans} that the equilibrium
distribution $w_0(n_+)$ exhibits non thermodynamical behaviour in
the sense that it becomes for large $n_c$ a non trivial function of
\begin{equation} \mylabel{c2}
x=\frac{n_+-n_-}{n_++n_-} \quad ,
\end{equation}
instead a Gaussean around the mean value of $x$ with a width 
$\propto 1/\sqrt{n_c}$. This solves a problem in the LMM and other
models which become trivial in the limit of many agents \cite{Ege}.
The transition probabilities \myref{c1} lead to a master equation.
In the limit of large $n_c$ it can be written in the form of a
Fokker-Planck equation for the distribution $w(x,t)$:
\begin{equation} \mylabel{c3}
\frac{\Delta w(x,t)}{\Delta t}=\frac{\partial}{\partial x}
    \left [ -A_x(x)\;w(x,t)+\frac{1}{2}\frac{\partial}{\partial x} 
    D_x(x)\;w(x,t))\right ]
\end{equation}
with a drift term
\begin{equation} \mylabel{c4}
A_x(x)=-2ax
\end{equation}
and a diffusion term
\begin{equation} \mylabel{c5}
D_x(x)=2b(1-x^2) \quad .
\end{equation}
The equilibrium distribution $w_0(x)$ depends only on the ratio
$\epsilon=a/b$ of the rate constants
\begin{equation} \mylabel{c6}
w_0(x)=\frac{(1-x^2)^{\epsilon -1}}{B(\epsilon ,1/2)}
\end{equation}
with the normalization given by Euler's $B$-function.
If a system is described by a Fokker-Planck equation, we can replace 
the original dynamic \myref{c1} for the individual agents by a
Langevin equation for $x$ 
\begin{equation} \mylabel{c7}
\frac{\Delta x}{\Delta t}=A_x+\sqrt{\frac{D_x}{\Delta t}}\; \eta_t
\end{equation}
with a white noise $\eta_t$. In equations \myref{c3} and \myref{c7}
no dependence on $n_c$ 
appears. It is hidden, however, by the condition, that the time
scale $\Delta t$ and the rates must satisfy
\begin{equation} \mylabel{c8}
a\Delta t,\; b\Delta t\;\propto \;1/n_c << 1 \quad ,
\end{equation}
if the dynamic \myref{c7} should represent the original 
dynamic \myref{c1}. In the numerical simulation we use \myref{c7} in
the form
\begin{equation} \mylabel{c9}
x(t+1)=x(t)-\epsilon \rho x + \sqrt{\rho(1-x^2)}\; \eta_t
\end{equation}
with an effective rate $\rho=2b\Delta t$. Using relation \myref{b2}
for the volatility, we arrive at a model similar to the one 
proposed by Alfarano and Lux \cite{AL}
\begin{equation} \mylabel{c10}
v(t)=v_0\left |x(t+1)-x(t) \right | \quad .
\end{equation}
Our version differs from \cite{AL} that in the latter the dynamic \myref{c1}
is used and the drift term is replaced by a reflecting boundary at $x^2=1$
as in the LMM. For small $x$ the system is described in both cases by
a random walk. Near $|x|=1$ the diffusion vanishes and the system 
remains a long time $T_w$ near the boundary. The average of 
the return time $T_w$ can be
computed analytically \cite{Hon}. For small $\epsilon$ we obtain
\begin{equation} \mylabel{c11}
T_w=\frac{1}{b\epsilon}\; +\;\frac{1}{b}\ln \frac{e}{2}\;+\;o(\epsilon) \quad .
\end{equation}
\begin{figure}[htb]
\let\picnaturalsize=N
\def\picsize{120mm}
\def\picfilename{fig6a.ps}
\ifx\nopictures Y\else{\ifx\epsfloaded Y\else\input epsf \fi
\let\epsfloaded=Y
\centerline{\ifx\picnaturalsize N\epsfxsize \picsize\fi
\epsfbox{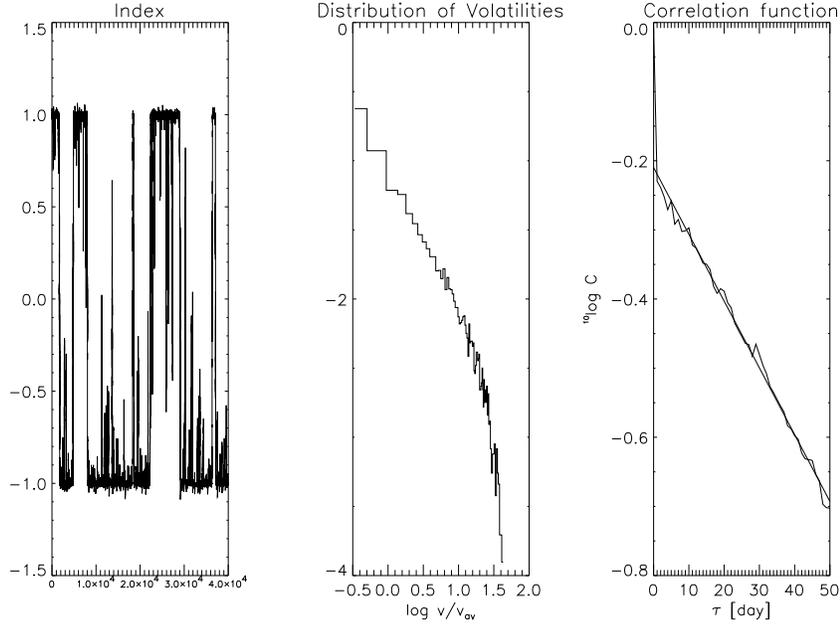}}}\fi
\medskip
\caption{\label{fig6} Simulation of the Kirman model. The left part shows the
         time serie $x(t)$, the middle part the volatility distribution
         and the right part the correlation function given by equ.
         \myref{d5}. The straight line
         corresponds to an exponential fit $\exp (-\gamma \tau )$ with
         $\gamma =0.022$.}
\end{figure}
Leaving after $T_w$ the boundary $x^2=1$, volatility clusters are formed by the
diffusion mechanism \myref{c9} with a time scale $1/b <<T_w$. 
In Fig.\ref{fig6} the results of a
simulation are shown with $\epsilon=0.02$ and $\rho =0.02$. The left part
gives $x(t)$ for $4\cdot 10^4$ time steps. The accumulation of points near
$|x|=1$ is clearly seen. In the middle part we show the distribution of the
volatility as function of $v/v_{av}$. The behaviour at large $v$ is compatible 
with the data, but not the peak at small $v$. On the right we show the 
correlation function of the volatility on log-linear scale. It agrees with 
the observed correlation (see Fig.\ref{fig5}), that a sharp drop at small
$\tau$ is followed by an exponential decay as indicated by the straight line
in Fig.\ref{fig6}. However, the drop is much to small and the decay
too strong as compared to the data. One could get agreement with $\epsilon$
values near 0.5, which would give totally unrealistic volatility
distributions.
Apart from the peak at small volatilities and the quantitatively
wrong correlations the Kirman model exhibits further
unwanted features. Since the diffusion is fast the value
of the volatilities in a cluster are always of the same size $v_0$ given 
by \myref{c10} and do not reproduce small size clustering seen in Fig.\ref{fig3}
and Fig.\ref{fig4}. The index values are bounded by
\begin{equation} \mylabel{c12}
|p-p_f|\le \frac{p_0}{n_f} \quad .
\end{equation}
Such a bound apart being unrealistic leads to a winning strategy for the 
fundamentalists in contradiction the efficient market hypothesis \cite{Fama}.
These deficiencies can be cured by using a three component model, where noise
traders can change into fundamentalists.

\section {Three component model} \label{three}
The three component model is obtained, if one includes in the agent 
dynamic \myref{c1} also the fundamentalists
$(i,k$ = $+,-,f)$. The system is described by the fraction $x$ as in equ.
\myref{c2} and the ratio of noise traders to fundamentalists
\begin{equation} \mylabel{t1 }
u=\frac{n_+ +n_-}{n_f} \quad .
\end{equation}
The total number $N=n_+ +n_- +n_f$ is to be kept constant. In order
to obtain a nontrivial distribution in $x,u$ in the limit of large 
$N$ we require the a symmetric herding matrix $b_{ik}$. The rate
parameters should be 
symmetric between optimistic and pessimistic noise traders leading to
\begin{equation} \mylabel{t2 }
b_{+k}=b_{+k}\; , \quad a_{+k}=a_{-k}\; ,\quad a_{k+}=a_{k-} \quad .  
\end{equation}
There remain only five independent rate constants in the
matrices $a$ and $b$. A last
condition comes from an integrability condition for the existence of a master
equation, as discussed by Aoki \cite{Aoki}. This reads in our notation
\begin{equation} \mylabel{t3 }
\frac{a_{+-}}{b_{+-}}=\frac{a_{f-}}{b_{f-}} \quad .
\end{equation}
Now we perform the same steps from the transition rates to the Langevin 
equation for $x$ and $u$ as in the Kirman model. The equation for $x$ is
the same as equ. \myref{c9}. For $u$ we obtain a second equation
\begin{equation} \mylabel{t4 }
u(t+1)=u(t)+A_u(u)\; +\; \sqrt{D_u(u)}\; \eta_t
\end{equation}
with a drift term
\begin{equation} \mylabel{t6 }
A_u(u)=(1+u)\left [(1-\epsilon_3/2)u+\epsilon\right ]\rho_3
\end{equation}
and a diffusion term
\begin{equation} \mylabel{t8 }
D_u(u)=\rho_3\ u \ (1+u)^2 \; ,
\end{equation}
where we introduced in analogy to \myref{c9} the effective rate
$\rho_3=2b_{+f}\Delta t$ and the ratio $\epsilon_3=a_{+f}/b_{+f}$. 
The expression for the volatility is given by
\begin{equation} \mylabel{t9 }
v(t)=v_0\left | u(t+1)x(t+1)-u(t)x(t) \right | \quad .
\end{equation}
The equilibrium distribution for the fractions $x$ and $u$
factorizes into a product
\begin{equation} \mylabel{t10}
w_0(x,u)=\frac{(1-x^2)^{\epsilon -1}}{B(\epsilon ,1/2)}\cdot
          \frac{u^{2\epsilon -1}(1+u)^{-(\epsilon_3+2\epsilon)}}
               {B(2\epsilon ,\epsilon _3)} \quad .
\end{equation}
In the two component model volatility clusters  are
due to a large return time for $\epsilon \to 0$. Unfortunately, small
$\epsilon$ are responsible for the
unwanted behaviour $\propto v^{-1}$ in the volatility distribution.
In the three component model one can avoid this drawback  
by a large time for a change from a small value $u_0$ to a large
value $u_1$  given by
\begin{equation} \mylabel{t11}
T(u_0 \to u_1)\propto (1+u_1)^{\epsilon_3-1} \; ,
\end{equation}
whereas the reversed time $T(u_1 \to u_0)$ is independent of $u_1$.
For simulations of the model
the parameters $\epsilon $ and $\epsilon_3$
can be determined from the equilibrium distribution of $v$.
Our choice $\epsilon =2$ and $\epsilon_3=6$ give a reasonable description
of volatility distribution as can be seen from Fig.\ref{fig1}.
With the rate parameters $\rho=0.001$ and $\rho_3=0.01$ the
empirical average waiting times shown in Fig.\ref{fig3} and the 
cluster parameter $\gamma$ shown in Fig.\ref{fig4} can be
reproduced. Especially the $u$ dependent return time \myref{t11} 
leads to clustering also for small $v$, as seen in Fig.\ref{fig4}.
With these values the correlation function shown in Fig.\ref{fig5}
can be predicted, which agrees with the data except in the transition
region between the drop and the slow decay.
Since we are using normalized volatilities, only the ratio
$\rho /\rho_3=0.1$ is important for waiting times and volatility
distributions. It means that changes in the
index are mainly caused by transitions between noise traders
and fundamentalists, and not by changes between optimists and
pessimists. The ratio $\epsilon_3/\epsilon =3$ implies that changes
from noise traders to fundamentalists are more influenced by the
spontaneous rate than the reversed process. Since the spontaneous
rate may represent private information, this seems not to be
unreasonable. From the average value of $<u>=0.4$ we find that
the fraction of noise traders is only roughly 30\%. 

\section {Conclusions} \label{concl}
The empirical data for a stock market as the DAX can be explained
by a model based on herding alone. Especially the time structure
of the volatility clustering can rule out a two component model. At
least three components are needed. Since the description of the data
is comparable or better than the predictions of the Lux-Marchesi model,
one may conclude that herding is the dominant mechanism in the LMM,
and the nonlinear coupling of the price evolution to the transition
probabilities is less important.\\
From the values of herding parameters ($\rho{_3}\sim 10\rho$) we find
the surprising fact that the herding effect is much stronger for
an agent in the fundamentalistic state than in the noise trader state.
A possible interpretation may be the following. A typical fundamentalistic
trader may be a fond manager. She follows the present majority if
herding dominates. In the opposite strategy not to follow the trend a
heavy risk for her career is buried. If the majority turns out to be
successful she will belong to the few who experience losses
 and will lose her job with
no opportunity to revise her strategy. This does not happen if she has
followed the majority.
On the long run agents not following the general trend will die out.
Therefore there exists a strong tendency towards herding behaviour
also for the fundamentalists.\\
{\bf Acknowlegments}: The author thanks Thomas Lux for stimulating discussions
and the numerical data of  a simulation of the LMM. He is grateful to
S.Alfarano for discussions and
Dr.\ Pierdzioch from the Istitut f\"ur Weltwirtschaft
for the access to the DAX data.

\end{document}